\documentstyle[preprint,aps,epsf,floats]{revtex}
\tighten

\def\btod{$\bar B \to D l \overline\nu$}
\def\btods{$\bar B \to D^* l \overline\nu$}
\def\bra#1{\langle #1 |}
\def\ket#1{| #1 \rangle}
\def\INSERTCAP#1#2{\vbox{%
{\narrower\noindent%
\multiply\baselineskip by 3%
\divide\baselineskip by 4%
{\rm Table #1. }{\sl #2 \medskip}}
}}
\def\np#1#2#3{\NP{\bf B#1} (#2) #3}
\def\pl#1#2#3{\PL {\bf B} {\bf #1} (#2) #3}
\def\prl#1#2#3{\PRL{\bf #1} (#2) #3}
\def\pr#1#2#3{\PR{\bf #1} (#2) #3}

\def\sjnp#1#2#3{\SJNP{\bf #1} (#2) #3}

\def\NP{{Nucl.\ Phys.\ }}
\def\PL{{Phys.\ Lett.\ }}
\def\PR{{Phys.\ Rev.\ }}

\def\PRL{{Phys.\ Rev.\ Lett.\ }}
\def\SJNP{{Sov.\ J. Nucl.\ Phys.\ }}

\begin{document}

\preprint{\vbox{
\hbox{CMU-HEP97-02}
\hbox{DOE/ER/40682-127}
\hbox{DOE/ER/41014-07-N97}
}}
\title{Analyticity, Shapes of Semileptonic Form Factors, 
and $\bar B \to \pi l \bar \nu$}
\author{C. Glenn Boyd\footnote{{\tt boyd@fermi.phys.cmu.edu}} 
}
\address{
Department of Physics, Carnegie Mellon University, 
\\
Pittsburgh, PA 15213
}
\author{Martin  J. Savage\footnote{{\tt 
savage@phys.washington.edu}}
}
\address{
Department of Physics, University of Washington, 
\\
Seattle, WA 
98195
}
\maketitle

\begin{abstract}

We give a pedagogical discussion of the physics underlying
dispersion relation-derived
parameterizations of form factors describing 
$B\rightarrow\pi l \bar \nu$ and $B\rightarrow D l \bar \nu$.
Moments of the dispersion relations are shown 
to provide substantially tighter constraints on the $f_+ (t)$ form factor 
describing $\bar B \to \pi l \overline \nu$
than the unweighted dispersion relation alone.
Heavy quark spin symmetry relations between the 
$B\rightarrow\pi l \bar \nu$ and $B^*\rightarrow\pi l \bar \nu$
form factors enables such constraints 
to be tightened even further.

\end{abstract}

\bigskip
\vskip 5.0cm
\leftline{February 1997}

\vfill\eject

\centerline{\bf Introduction}
\bigskip
\bigskip

Exclusive semileptonic decays of heavy mesons play an
important role in the determination and over-constraining 
of the Cabibbo-Kobayashi-Maskawa mixing matrix. The 
CKM element $V_{cb}$ has been extracted\cite{BtoD} from
\btods\ and \btod\ using
heavy quark symmetry\cite{hqs}, while the element
$V_{ub}$ has been estimated from
$\bar B \to \pi l \overline \nu$ and 
$\bar B \to \rho l \overline \nu$ rates~\cite{pirho}
using various models. 
In both cases, the normalization and
shape of the relevant hadronic form factors influence the
extracted value of the CKM angle. 
For $V_{cb}$,
the normalization of the $B\rightarrow D^{(*)}$
matrix element at zero recoil is provided by heavy 
quark symmetry.
However, typical extrapolations to this point
use ad-hoc  parameterizations of form factors that
introduce theoretical uncertainties comparable to 
the statistical uncertainties\cite{bgl2,stone}.  
This is
especially unfortunate since the uncertainty 
in $V_{cb}$ feeds into
unitarity-triangle constraints from  CP violation observed 
in the kaon system as the fourth power\cite{utri}. 
For $V_{ub}$ neither the normalization nor the shape is 
well known.  The normalization near zero recoil
may be obtained from lattice simulations or by combining
heavy quark and chiral symmetries with measurements of
related semileptonic decays in the charmed and bottom 
sector\cite{hqchiralsym},
but a parameterization away from zero recoil is necessary to
compare to experimental data.

Some progress in describing the shape of such form factors
has recently been made in the form of model-independent 
parameterizations\cite{bgl2,bl} based on QCD dispersion
relations and analyticity\cite{hist,brm}. These dispersion relations  
lead to an infinite tower of upper and lower
bounds that can be derived by using 
the normalizations of the form factor $F(t_i)$
at a fixed number of kinematic points $t_i$ as 
input\cite{brm,bgl1,lellouch}. When the normalization is known
at several points (say, five or more for 
$\bar B \to \pi l \overline \nu$),  the upper and lower bounds
are typically so tight they look like a single line. 
A natural question
then arises: 
What is the most general  form consistent 
with the constraints from QCD?
The answer to this question is the parameterization of 
reference\cite{bgl2}.  For a generic form factor $F(t)$ describing
the exclusive semileptonic decay of a $\bar B$ meson to a final
state meson $H$ as a function of momentum-transfer squared 
$t = (p_B -p_H)^2$, 
the parameterization takes the form 
\begin{eqnarray}\label{master}
F(t) = {1\over P(t) \phi(t)} \sum_{k=0}^\infty a_k\  z(t;t_0)^k 
\ \ \ \ ,
\end{eqnarray}
where $\phi(t)$ is a computable function arising from 
perturbative QCD.  The function $P(t)$ depends only on
the masses of mesons below the $\bar B\bar H$ 
pair-production threshold
that contribute to $\bar B\bar H$ pair-production as virtual
intermediate states.
The variable $z(t;t_0)$ is a kinematic function of $t$ defined by
\begin{eqnarray}\label{zdef}
{1+z(t;t_0) \over 1-z(t;t_0)} = \sqrt{ t_+ -t \over t_+ - t_0}
\ \ \ \ \ ,
\end{eqnarray}
where $t_+ = (M_B + M_H)^2$ is the pair-production threshold
and $t_0$ is a free parameter that is 
often\cite{bgl2,hist,brm,bgl1,lellouch} taken to be
$t_- = (M_B - M_H)^2$, the maximum momentum-transfer squared
allowed in the semileptonic decay $\bar B \to H l \overline \nu$.
The coefficients $a_k$ are unknown constants constrained to obey
\begin{eqnarray}\label{asum}
\sum_{k=0}^\infty \left( a_k \right)^2 \leq 1
\ \ \ \  .
\end{eqnarray}
The kinematic function $z(t;t_0)$ takes its minimal 
physical value $z_{min}$ 
at $t=t_-$, vanishes at $t=t_0$, and reaches its
maximum $z_{max}$ at $t=0$.
Thus the sum $\sum a_k\ z^k$ is a series expansion about 
the kinematic point $t=t_0$.
For \btods\ with 
$t_0 = t_-$, the maximum value of $z$ is $z_{max} = 0.06 $, and the
series in Eq.~(\ref{master})\ can be truncated while introducing
only a small error~\cite{bgl2}. The value $z_{max}$
can be made even smaller by choosing an
optimized value $0 \le t_0 \le t_-$\cite{bl}.
In that case, most form factors describing \btod\
and \btods\ can be parameterized with only one unknown constant
to an accuracy of a few percent (assuming the normalization at zero 
recoil given by heavy quark symmetry).
Thus the continuous function $F(t)$ has been reduced to a single
constant, for example the value of the form factor $F(t=0)$
at maximum recoil. For $\bar B \to \pi l \overline \nu$, the maximum
value of $z$ is $z_{max} = 0.52$, but even in this case 
Eqs.~(\ref{master})\
and (\ref{asum})\
severely constrain the relevant form factor\cite{bgl1,lellouch}. 
 
This remarkable constraining power can
be traced to the existence of a naturally small parameter $z_{max}$
that arises algebraically from a conformal map. 
In this paper we attempt to trace the physical origin of 
$z_{max}$ in the hope of developing 
some intuition about the physics underlying 
the analyticity constraints of Eqs.~(\ref{master})\
and (\ref{asum}).
Further, we will incorporate
two generalizations that lead to a significantly stronger constraint
on the observable $\bar B \to \pi l \overline \nu$ form factor.

\bigskip
\bigskip
\centerline{\bf Physical Basis for a Small Parameter}
\bigskip
\bigskip

To understand heuristically why there is a small parameter 
associated with semileptonic heavy meson decays, 
consider for the moment a form factor $F(t)$ in the decay \btod,
and take $t_0=t_-$.  Crossing symmetry tells us the analytic continuation of 
the form factor $F(t)$ that describes semileptonic
decay for $0 \le t \le t_-$ also describes $\bar B\bar D$ pair
production for $t \ge t_+$. Fig.~1 shows the general features
one expects for $F(t)$ in the region $0 \le t \le \infty$.
The form factor has a cut due to pair-production
beginning at $t = t_+$, as well as a series of 
poles from bound $B_c$-type states in the vicinity of $t_+$.
It varies rapidly near these poles, then falls smoothly 
from its peak values near $t \sim t_+$ to its minimum values
near $t \sim 0$. It is not essential to our argument that
the form factor decreases monotonically as $t$ approaches zero,
only that the variation in $F(t)$ over the semileptonic 
region $0 \le t \le t_-$ is determined by the distance
to the branch cut $t = t_+$ and the magnitude of the form factor
near the branch point, $F(t_+)$.  
For fixed $F(t_+)$, $F(t)$ varies more slowly over the semileptonic
region as $t_-/t_+$ decreases, while for fixed $t_-/t_+$, $F(t)$
also varies more slowly in the semileptonic region as $F(t_+)$ decreases.
Both observation and QCD perturbation theory imply that 
the rate of $\bar B\bar D$ pair-production cannot be
arbitrarily large for $t>t_+$, and combined with the fact that 
$t_+ \gg t_-$ for \btod, we
expect the variation of $F(t)$ over the semileptonic region
to be small. 
We wish to associate the small parameter $z_{max}$
with this variation.

\begin{figure}
\epsfxsize=13cm
\hfil\epsfbox{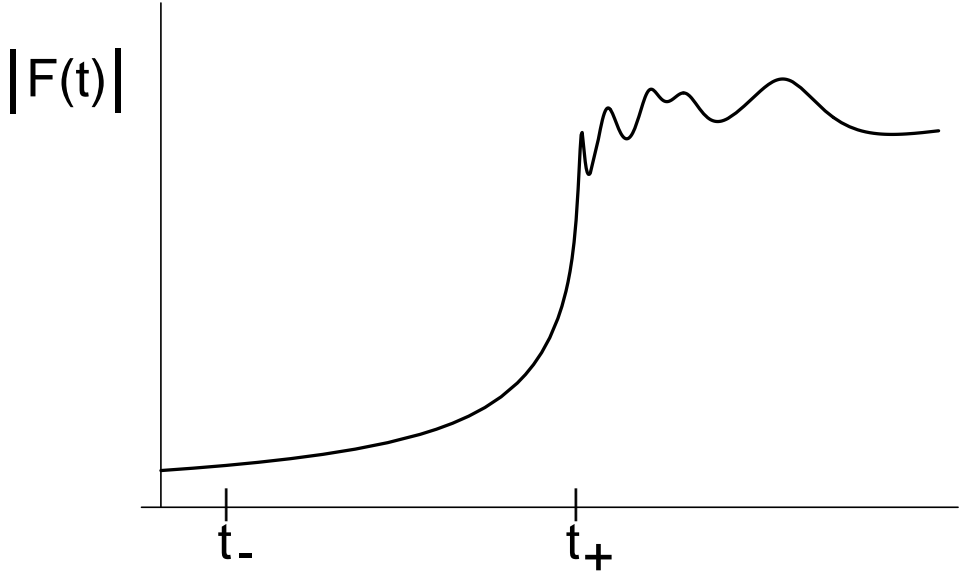}\hfill
\caption{
The magnitude of a generic form factor $F(t)$ as a function
of $t$.
The pair-production threshold $t_+$ and the semileptonic endpoint
$t_-$ are shown schematically.
}
\label{tplane}
\end{figure}

Suppose the form factor can be roughly described in the
physical semileptonic region by 
\begin{eqnarray}\label{ppole}
F(t) \sim {F_0 \over (t_+ -t)^p}
\ \ \ \ \ ,
\end{eqnarray}
where $F_0$ is a constant. 
A reasonable measure of the variation of $F(t)$ over the
physical region for semileptonic decay is
\begin{eqnarray}\label{rparam}
\delta_F & = & { F(t_-) - F(0) \over F(t_-) +  F(0) }
\nonumber\\
 & = & { t_+^p - (t_+ - t_-)^p \over t_+^p + (t_+ - t_-)^p }\ \ .
\end{eqnarray}
This measure depends only on the kinematic thresholds $t_+,t_-$
and the power $p$. 
For $p=1$ the form factor is pole dominated and  
$\delta_F$ is similar to the Shifman-Voloshin parameter\cite{SV},
$( M_B - M_D)^2/ (M_B + M_D)^2 \sim 1/4$.  
However, comparison with Eq.~(\ref{zdef})\ reveals that $\delta_F$ can be
identified with $z_{max}$ only if $p= 1/2$ giving
\begin{eqnarray}
\delta_F & = & \left( {\sqrt{M_B}-\sqrt{M_H}\over 
\sqrt{M_B}+\sqrt{M_H} }\right)^2
 = z_{max}
\ \ \ \ .
\end{eqnarray}
This value
of $p$ leads to the small value of $z_{max}$ for $B\rightarrow D$.
Other decays such as $\bar B \to \rho l \overline \nu$, 
$D \to K^* l \overline \nu$, etc., have larger values of $z_{max}$,
with the largest  occurring for $\bar B \to \pi l \overline \nu$.
Even for this extreme case, $z_{max} \approx 1/2$ is small enough
to provide a useful expansion parameter.

On the face of it, this value of $p$ seems rather surprising. After
all, we know bound states exist and will contribute to form factors
like poles. On the other hand, the dispersion relation relies on
quark-hadron duality and perturbative QCD. 
In perturbative QCD the
fundamental degrees of freedom are quarks and gluons, there
are no bound states at any finite order in perturbation theory 
to couple to the pair-produced fermions, and the form factor has
no poles.  Indeed, at leading order in the Parton Model, the
\btods\ form factors have the form \cite{suzuki}\
of Eq.~\ref{ppole},
\begin{eqnarray}
F(t) = {F_0 \over \sqrt{t_+ -t} }
\ \ \ \ ,
\end{eqnarray}
with $p=1/2$.
Given that there are bound states in nature, how can the perturbative 
QCD results be trustworthy? 
Certainly perturbative QCD cannot
be used directly in the semileptonic region. 
However, the
perturbative calculation
of pair-production should be reliable as long as a large region of
momentum transfer is
smeared over \cite{poggio}, or 
integrated over with smooth weighting functions.
By constraining the magnitude
of the form factor in the pair-production region, the perturbative
analysis indirectly constrains the shape of the form factor in the
semileptonic region.

For $\bar B \to \pi l \bar \nu$, the kinematically allowed region
$t_-$ is much larger, and the heuristic discussion above applies
less clearly. An explicit derivation is required to see that,
even in this case, pair-production constraints 
allow $F(t)$ in the semileptonic region to be expanded in
powers of $z \le z_{max}$.

\bigskip
\bigskip
\centerline{\bf Moments of the Dispersion Relation}
\bigskip
\bigskip

For a general semileptonic decay $\bar B \to H l \overline \nu$,
the heuristic discussion of the previous section can be made concrete
by considering the two-point function for the vector or axial vector currents
$J= \bar q \gamma^\mu b, \ \bar q \gamma^\mu \gamma_5 b$ that 
arise in the charged 
current decay of b-hadrons,
\begin{eqnarray}\label{pimunu}
\Pi_J^{\mu \nu}(q) & = &   {i \int d^4\!x \, e^{iqx}\langle 0|{\rm T}
J^\mu(x) J^{\dagger\nu}(0)|0 \rangle}
\cr
& = & 
(q^\mu q^\nu-q^2g^{\mu\nu})\Pi_J^T(q^2) +
g^{\mu\nu}\Pi_J^L(q^2)
\ \ \  .
\end{eqnarray}
The polarization functions $\Pi^{L,T}_J (q^2)$ do not fall off fast 
enough at large $q^2$ for an 
unsubtracted dispersion relation to be finite.
However, derivatives of the polarization functions do fall fast 
enough at high $q^2$ for finite dispersion relations to exist.
As we wish to constrain hadronic form factors by a perturbative 
calculation, it is useful to define the derivatives of the polarization 
functions at $q^2=0$
where the partonic amplitude is well behaved, far from the 
physical region for
$\bar B\bar H$ pair production.  
At $q^2=0$ the $n^{th}$-derivative of the 
polarization tensor for
the current $J^\mu$ is
\begin{eqnarray}\label{ndisp}
\chi^{(n)}_J &\equiv & 
{1\over 3 \Gamma(n+3)} {\partial^{n+2}\Pi^{ii}_J(0)\over 
\partial (q^2)^{n+2} }
\nonumber\\
& = & 
{1 \over 3 \Gamma(n+2)}   {\partial^{n+1}\Pi^T_J(0)
\over \partial (q^2)^{n+1} }
- {1 \over 3 \Gamma(n+3)} { \partial^{n+2}\Pi^L_J(0)
\over \partial (q^2)^{n+2} }
\nonumber\\
 &=& {1\over\pi}\int_0^\infty dt \, {
\frac13 {\rm Im}\,\Pi^{ii}_J(t) \over t^{n+3}  }
\ \ \  ,
\end{eqnarray}
where $J=V, A$ for vector and axial vector currents respectively.
This dispersion relation relates the computation of
$\chi_J^{(n)}$ at the unphysical value $q^2=0$ to
the weighted integral over the pair-production region
of the imaginary part of $\Pi^{ii}_J (q^2)$.
The higher the moment $n$ the more the integral is weighted 
near the pair production threshold, so
we expect the calculation to be most 
reliable for low moments where the smearing is largest.

It is straightforward to determine the $\chi_J^{(n)}$ in perturbative 
QCD.  For a 
ratio of quark masses $u = m_q/m_b$, a one-loop
(leading order) calculation of the vector current correlator gives
\begin{eqnarray}\label{chiF}
\chi^{(n)}_V (u) & = & { 3 \left(\Gamma (n+3) \right)^2 \over 
    2 \pi^2 m_b^{2 n+2} \Gamma (2 n +6) }
\Biggl[ {1\over n+1} F(n+1, n+3; 2 n +6; 1- u^2)
\nonumber \\
&+& {1 -u\over 4 (n+2) } \biggl[ u F(n+2, n+4; 2 n+7; 1- u^2)
 - F(n+2, n+3; 2 n+7; 1-u^2) \biggr] \Biggr],
\end{eqnarray}
where $F(a,b;c;\xi)$ is a hypergeometric function. 
The same
expression results for the  axial current after the substitution
$u \to -u$, i.e.  
$\chi_A^{(n)} (u) = \chi_V^{(n)} (-u)$.
For a massless quark $m_q=0$, the expressions
simplify to
\begin{eqnarray}
\chi^{(n)}_{V,A}(0) = { 3 \over 4 \pi^2 m_b^{2 n+2} }
{1 \over (n+1)(n+2)(n+4)}
\ \ \ \ .
\end{eqnarray}
The correlator Eq.~(\ref{pimunu})\ 
has also been computed at two loops\cite{generalis}, i.e.
${\cal O}\left(\alpha_s\right)$.
The higher order corrections result in 
a $25\%$ increase\cite{lellouch} in $\chi^{(0)}(0)$.

Since production of $\bar B \bar H$ hadrons is a subset of 
total hadronic production, the perturbative calculation $\chi^{(n)}$
serves to constrain the analytically continued form factors
for $\bar B\rightarrow H$ decay.  
More precisely, the partonic 
computation provides an upper bound to the smeared contributions 
of poles and cuts above the pair production threshold. The 
contribution of poles below threshold will also 
influence the variation of a given form factor $F(t)$ in
the semileptonic region, and must be considered separately. 
While sub-threshold contributions are not a fundamental aspect of 
the dispersion
relation approach (for example, form factors in $D \to \pi
l \overline\nu$ are analytic below the $D-\pi$ threshold),
they are not accounted for by the perturbative calculation
and must be properly handled when present\cite{spoilers}.

We now turn to relating $\chi_J^{(n)}$ to $F(t)$.
This is accomplished by inserting a sum over intermediate
states into ${\rm Im}\,\Pi^{ii}_J (q^2)$,
\begin{eqnarray}\label{optical}
{\rm Im}\,\Pi^{ii}_J(q^2) = 
  {1\over 2}\int {d^3p_1 d^3p_2 \over (2 \pi)^2 4 E_1 E_2 }
  \delta^{(4)}(q-p_1-p_2) \sum_{pol}
 \langle 0 | J^{\dagger i}  |\bar B(p_1) \bar H(p_2)\rangle
\langle \bar B(p_1) \bar H(p_2) | J_i |0 \rangle + \ldots \, ,
\end{eqnarray}
where the sum is over polarizations of $H$ and the ellipsis 
denotes strictly positive contributions from the $B^*$, 
higher resonances and multi-particle states. In terms
of a calculable kinematic function $k(t)$ arising from 
two-body phase space and the Lorentz structure associated with the 
form factor $F(t)$, 
we may substitute the inequality
\begin{eqnarray}
\frac13 {\rm Im}\,\Pi^{ii}_J(t) \ge k(t) |F(t)|^2
\ \ \  ,
\end{eqnarray}
into Eq.~(\ref{ndisp})\ 
to get the contribution to the hadronic moment $\chi_J^{(n)} (hadronic)$ from the 
form factor $F(t)$ of interest,
\begin{eqnarray}
\chi_J^{(n)} (hadronic) & \geq & {n_I\over \pi} \int_{t_+}^\infty 
{k(t)\ |F(t)|^2 \over t^{n+3} } 
\ \ \ \ ,
\end{eqnarray}
where $n_I$ is the isospin degeneracy of the $\bar B \bar H$ pair.
We rely on perturbative QCD at the unphysical point $q^2=0$ (or
equivalently, on global duality for suitably smeared
production rates) to assert that hadronic
and partonic expressions for $\chi$ are equal.
Then for each $n$,
\begin{eqnarray}
\chi_J^{(n)} (hadronic)\ = \ \chi_J^{(n)} (u)
\ \ \ ,
\end{eqnarray}
where $\chi_J^{(n)} (u)$ is the $n^{th}$ moment as computed in perturbative QCD.
Therefore we have that
\begin{eqnarray}
 {n_I\over \pi \chi_J^{(n)} (u)} \int_{t_+}^\infty  {k(t)\ |F(t)|^2 \over t^{n+3} } 
 & \leq & 1
\ \ \ \ ,
\end{eqnarray}
and hence
\begin{eqnarray}\label{tint}
{1\over \pi} \int_{t_+}^\infty dt | h^{(n)}(t) F(t) |^2 \le 1
\ \ \ \  ,
\end{eqnarray}
where $ h^{(n)}(t) = \sqrt{ n_I k(t) \over \chi_J^{(n)} (u)}\  t^{-(3+n)/2} $. 
The argument of the
square root is positive since the integrand came
from a production rate.

The inequality of 
Eq.~(\ref{tint})\ makes  clear how the perturbative calculation
constrains the magnitude of the form factor in the
pair-production region. 
To constrain the form factor in
the semileptonic region $0 \le t \le t_-$, we would like
to find functions $\varphi_k (t)$ that are orthonormal with
respect to the integral Eq.~(\ref{tint}),
\begin{eqnarray}\label{orthon}
{1\over \pi} \int_{t_+}^\infty dt \, {\cal R}{\rm e} 
   \left[ \varphi_k (t)\ \varphi_j^* (t) \right]\ 
& = &   \delta_{k j} 
\ \ \ \  ,
\end{eqnarray}
and that vanish somewhere in the semileptonic region, say at 
$0 \le t_0 \le t_-$. 
We could then expand $h^{(n)}(t) F(t)$
in terms of these basis functions and use Eq.~(\ref{tint})\ to
bound the expansion coefficients
\footnote{
We can choose the expansion coefficients $a_k$ to be real
so that only 
${\cal R}{\rm e } [ \varphi_k (t)\ \varphi_j^* (t)]$ need
vanish for $k \ne j$ since it is this expression that 
arises in 
$ | \sum a_k \varphi_k|^2$. 
In a more general case where
the $a_k$ are complex, our results go through unchanged if
the inner product in Eq.~(\ref{orthon})\ is redefined as 
$ <f,g> \equiv \lim_{\epsilon \to 0}
{1\over 2 \pi} \int_{t_+}^\infty 
dt \, [ f(t+i\epsilon) g^*(t+i\epsilon) + 
f(t-i\epsilon) g^*(t-i\epsilon) ] .$}. 
If $h^{(n)}(t) F(t)$ turned out
to be analytic in $t$ outside the pair production region,
its expansion would be equally valid 
in the semileptonic region, and we would have a parameterization
of $F(t)$ in terms of unknown, but bounded, expansion coefficients.
Unfortunately, neither $h^{(n)}(t)$ nor $F(t)$ are in general analytic
away from the pair production cut. The kinematic
factor $h^{(n)}(t)$ has explicit poles at $t=0$ and the form factor 
$F(t)$ may also have poles arising from the contribution
of bound states that can interpolate between the current $J$
and the $\bar B\bar H$ pair. For example, the experimentally
accessible form factor $f_+(t)$ in $\bar B \to \pi l \overline\nu$\ has
a pole at $t = M_{B^*}$ coming from the contribution of the
$B^*$ resonance.

Fortunately, a simple pole at $t= t_p$ can be eliminated by 
multiplying by $z(t;t_p)$. Rewriting $z$ (as defined in 
Eq.~(\ref{zdef})\ )
as
\begin{eqnarray}\label{blaschke}
z(t; t_p) & = &  
{ t_p - t \over (\sqrt{t_+ -t} + \sqrt{t_+ - t_p} )^2} 
\ \ \ \ ,
\end{eqnarray}
makes it clear
that $z(t;t_p)$ vanishes at $t=t_p$ and has magnitude one in the 
pair-production region, $|z(t; t_p)|=1$ for $t \ge t_+$. We can
therefore construct a quantity with no poles outside the 
pair-production region by multiplying $h^{(n)}(t)$ and $F(t)$ by
factors of $z(t;t_p)$ for each pole at $t_p$. To make 
an analytic function outside the pair-production region, 
we generally also need to eliminate square-root
branch cuts in $h^{(n)}(t)$ that arise from factors of
the $H$-meson three-momentum by dividing by $\sqrt{z(t,t_-)}$.
The elimination of poles and cuts from $h^{(n)}(t)$ by
$h^{(n)}(t) \to \tilde P(t) h^{(n)}(t)$,
where $\tilde P(t)$ is a product of $z(t;0)$'s and
$\sqrt{z(t,t_-)}$'s,  can be automatically
accomplished by replacing 
\begin{eqnarray}\label{bkaschkeizing}
{1\over t} & \to & { - z(t;0) \over t} 
         = {1 \over (\sqrt{t_+ -t} + \sqrt{t_+})^2 }
\nonumber\\
\sqrt{t_- - t} & \to & \sqrt{  t_- -t\over z(t;t_-)}
=  \sqrt{t_+ -t} + \sqrt{t_+ - t_-} 
\ \ \ \  .
\end{eqnarray}
The elimination of poles from $F(t)$ by  $F(t) \to P(t) F(t)$ is
accomplished by multiplying by a product $P(t) = \Pi_j z(t;t_j)$
for each contributing sub-threshold resonance
of invariant mass-squared $t_j$. Since $ \tilde P(t)$ 
and $P(t)$ have unit modulus along the pair production cut 
(the integration region in  Eq.~(\ref{tint})), the  well-behaved
quantity $ \tilde P(t) h^{(n)}(t) P(t) F(t)$ obeys the same relation,
\begin{eqnarray}\label{Ptint}
{1\over \pi}\int_{t_+}^\infty dt |\tilde P(t) h^{(n)}(t) P(t) F(t)|^2 \le 1
\ \ \ \  .
\end{eqnarray}
Whereas $\tilde P(t)$ may be viewed as a technical device to
smooth out the kinematic function $h^{(n)}(t)$, $P(t)$ contains
essential information about the resonance structure of $F(t)$
in the unphysical region $ t_- < t <  t_+$.  In Eq.~(\ref{tint}),
the poles in $F(t)$ above threshold are constrained by the
perturbative calculation; in Eq.~(\ref{Ptint}), the poles
below threshold are accommodated by $P(t)$. 
Both sets of poles
influence the shape of $F(t)$ in the semileptonic region. 
Since
$P(t)$ depends only on the position of the poles below threshold
and not on the residues, it applies for arbitrarily strong or weak
residues.
Therefore, we should not be surprised that
the eventual effect of a non-trivial function $P(t)$
is to weaken the constraint on $F(t)$.

In terms of orthonormal functions $\varphi_k (t) $ satisfying
Eq.~(\ref{orthon}), the expansion
\begin{eqnarray}\label{taylor}
\tilde P(t) h^{(n)}(t) P(t) F(t) = \sum_{k=0}^\infty a^{(n)}_k \varphi_k (t)
\ \ \ \ 
\end{eqnarray}
combines with Eq.~(\ref{Ptint})\ to yield
\begin{eqnarray}\label{ansum}
\sum_{k=0}^\infty \left( a^{(n)}_k\right) ^2 \le 1
\ \ \ \ ,
\end{eqnarray}
valid for moderate values of $n \ge 0$. 
As the expression given in Eq.~(\ref{taylor})\ 
is valid everywhere outside the cut in the complex $t$ plane the form factor
$F(t)$ in the region of semileptonic decay $0 \le t \le t_-$ is
\begin{eqnarray}\label{funnyparam}
F(t) = {1 \over \tilde P(t) h^{(n)}(t) P(t)} \sum_{k=0}^\infty a^{(n)}_k 
\varphi_k (t)
\ \ \ .
\end{eqnarray}

The sum is over
$k \ge 0$ because by construction  $\tilde P(t) h^{(n)}(t) P(t) F(t)$
has no poles for $t < t_+\ .$
All that remains is to find the orthogonal polynomials $\varphi_k (t)$. 
This 
a math problem that can be accomplished by a change
of variables. 
In the complex $t$ plane, the integration contour
may be viewed as a segment 
from $+\infty$ to $t=t_+$ just below the cut and a segment 
from $t=t_+$ to $+\infty$ just above the cut. Defining
$y = \sqrt{ t -t _+}$ maps the line segments just above and
below this cut onto the real $y$ axis. The $y$ axis in turn
can be mapped onto the unit circle by the bilinear transformation
$z(t;t_0) = (y - \sqrt{t_0 -t_+} )/(y + \sqrt{t_0 -t_+} )$. This is 
precisely the change of variables in Eq.~(\ref{zdef}). Since
$z^n= e^{i n\theta}$ are orthonormal functions on the unit circle, we can
work backwards to find
\begin{eqnarray}\label{varphik}
\varphi_k (t)\equiv 
{1\over \sqrt{t_+ -t} + \sqrt{t_+ -t_0} }
    \Biggl( {t_+ -t_0 \over t_+ -t}\Biggr)^{1/4}
 \Biggl( { \sqrt{t_+ -t} - \sqrt{t_+ -t_0} 
     \over \sqrt{t_+ -t} + \sqrt{t_+ -t_0} } \Biggr)^k 
\ \ \ \ \ .
\end{eqnarray}
The expansion in orthonormal basis functions is simply a Taylor 
series in $z(t;t_0)^k$. However, the variable $z$ 
does little to aid the development of physical intuition,
so we continue to work with the momentum transfer 
$t$.  
Contact with previous literature\cite{bgl2,bl}
can be made by identifying 
\begin{eqnarray}\label{phidef}
\phi^{(n)}(z(t;t_0)) =  
\left(\sqrt{t_+ -t} + \sqrt{t_+ -t_0}\right)
     \Biggl( {t_+ -t \over t_+ -t_0}\Biggr)^{1/4}
    \tilde P(t) h^{(n)}(t) 
\ \ \ \  ,
\end{eqnarray}
choosing $n=0$, setting $t_0 = t_-$ (or, in the case
and language of reference\cite{bl},  
$t_0= (1-N) t_+ + N t_-$ ), and expressing 
the ``Blaschke factors''\cite{cap,rt2,duran} $z(t;t_p)$
composing $P(t)$ in terms of $z(t;t_-)$ and $z(t_p; t_-)$. 
With these identifications
Eq.~(\ref{funnyparam})\ becomes Eq.~(\ref{master})\
with $\phi(z) = \phi^{(0)}(z)$ and
$a_k = a_k^{(0)}$.

\bigskip
\bigskip
\centerline{\bf Parameterizations for Semileptonic Form Factors}
\bigskip
\bigskip

We are primarily interested in constraining form factors that
describe the decay of B mesons, although the formalism applies
equally well to  $\Lambda_b$ baryons, or even $D$ and $K$ mesons
if $\Pi(q^2)$ is evaluated at an appropriate spacelike $q^2$. 
Specialization to a particular decay and form factor requires
an explicit computation of the $\phi$ functions. For a pseudo-scalar
final meson $H$ or a vector meson $H^*$ with polarization
$\epsilon$, the various form factors in semileptonic $B$ decay
may be defined by
\begin{eqnarray}\label{Fdef}
  \bra{H^*(p',\epsilon)} V^\mu \ket{\bar B(p)} &=&
    \, i g \epsilon^{\mu \alpha \beta \gamma} \epsilon_\alpha^* \,
                          p_{\beta}' \, p_{\gamma} 
\nonumber\\
  \bra{H^*(p',\epsilon)} A^\mu \ket{\bar B(p)} &=&
   \, f \epsilon^{*\mu} + (\epsilon^{*} \! \cdot p)
   [a_+ (p + p')^\mu + a_- (p - p')^\mu] \cr
\nonumber\\
   \bra{H(p')} V^\mu \ket{\bar B(p)} &=& \, f_+ (p + p')^\mu + f_-
   (p - p')^\mu
\ \ \ \ ,
\end{eqnarray}
where it is useful to also define
\begin{eqnarray}
F_1= {1\over M_H} \left[\frac12 (t_+ -t)(t_- -t) a_+ 
- \frac12 (t-M_B^2+M_H^2)f
\right] 
\ \ \ \ ,
\end{eqnarray}
with $t = (p-p^\prime)^2$.
It is straightforward to determine the $k(t)$ function 
associated with each of the 
form factors,
\begin{eqnarray}
k_i (t) & = & {1\over 3\pi\ 2^s} \left( {1\over t} \right)^p
\left[(t-t_+)(t-t_-)\right]^{w/2}
\ \ \ \  .
\end{eqnarray}
For the form factors whose contribution to the rate is unsuppressed
by the lepton mass, the indices $s,p$ and $w$ are given by
\begin{eqnarray}
k_g (t) &:& s=5 \ ;\ p=1\ ;\ w=3
\nonumber\\
k_{F_1} (t) &:& s=4 \ ;\ p=2\ ;\ w=1
\nonumber\\
k_{f} (t) &:& s=3 \ ;\ p=1\ ;\ w=1
\nonumber\\
k_{f_+} (t) &:& s=4 \ ;\ p=2\ ;\ w=3
\ \ \ \  .
\end{eqnarray}
The $\phi^{(n)} (t)$ functions defined in Eq.~(\ref{phidef})\ for 
each form factor are
\begin{eqnarray}\label{Phi}
\phi_i^{(n)} (t;t_0) &=& \sqrt{ n_I\over 2^s 3 \pi \chi_J^{(n)} }
\left( {t_+-t\over t_+-t_0}\right)^{1/4}
\left(\sqrt{t_+-t} + \sqrt{t_+}\right)^{-(3+n+p)}
\nonumber\\
&&(\sqrt{t_+ -t} + \sqrt{t_+ -t_0})
(\sqrt{t_+ -t} + \sqrt{t_+ -t_-})^{w/2}
(t_+ - t)^{w/4}
\ \ \ \ .
\end{eqnarray}
Our previous discussions allow us to see that
each particular form factor $F_i$ has 
the functional form
\begin{eqnarray}\label{nmaster}
F_i(t) = {1\over P_i(t) \phi_i^{(n)}(t;t_0)} 
\sum_{k=0}^\infty a^{(n)}_k z(t;t_0)^k 
\ \ \  ,
\end{eqnarray}
for each moment $n$ and expansion point $t_0$,
where $z$ may be expressed as
\begin{eqnarray}\label{z0def}
z(t;t_0) \equiv { \sqrt{t_+ -t} - \sqrt{t_+ -t_0} \over
     \sqrt{t_+ -t} + \sqrt{t_+ -t_0}  }
     \ \ \ \ ,
\end{eqnarray}
and
$\sum (a_k^{(n)})^2 \le 1$.

The Blaschke factors $P_i(t)$ depend on the masses of sub-threshold
resonances. For \btod\ and \btods\ from factors, the masses
of the relevant $B_c$-type resonances can be rather accurately
estimated from potential models\cite{quigg,klt}.
Using the results of reference\cite{quigg} in Eq.~(\ref{blaschke}), 
the Blaschke factors for the form factors $f$ and $F_1$ are
\begin{eqnarray}
P_{f}(t) &=& P_{F_1}(t) 
\nonumber\\
&=& z(t; ( 6.730\, GeV)^2 ) z(t;(6.736\, GeV)^2)
z(t;(7.135\,  GeV)^2) z(t; (7.142\,  GeV)^2) 
\ \ \ \ ,
\end{eqnarray}
while for the form factors $g$ and $f_+$ they are 
\begin{eqnarray}
P_{g}(t)  = z(t;(6.337\, GeV)^2) z(t; (6.899\,  GeV)^2)
z(t;(7.012\, GeV)^2) z(t;(7.280\, GeV)^2)
\ \ \  ,
\end{eqnarray}
and 
\begin{eqnarray}
 P_{f_+}(t) = z(t;(6.337\, GeV)^2) z(t; (6.899\,GeV)^2)
       z(t;(7.012\, GeV)^2)
\ \ \ ,
\end{eqnarray}
respectively.  For $\bar B \to \pi l \overline \nu$, there
is only one resonance and the form factor $f_+ (t)$ has
the Blaschke factor $P(t) = z(t; (5.325\, GeV)^2)$. For
$D \to \pi l \overline\nu$, there are no resonances, and  
$P(t) =1$.

For the lowest moment ($n=0$)  and 
$t_0 = t_-$, Eqs.~(\ref{Phi}) - (\ref{z0def})
reproduce the results of references\cite{bgl2}, while for arbitrary
$t_0$ they reproduce the results for mesons given in \cite{bl}.
For higher moments ($n > 0$) and a given form factor $F_i$,
they imply 
\begin{eqnarray}\label{cksum}
\sum_{k=0}^\infty a^{(n)}_k z (t;t_0)^k &=&
    {\phi_i^{(n)}(t) \over \phi_i^{(0)}(t)} 
    \sum_{k=0}^\infty a^{(0)}_k z(t;t_0)^k
\nonumber\\
 &=& 
\sqrt{ \chi^{(0)}_J \over \chi^{(n)}_J }
\Biggl[ {1 \over \sqrt{t_+-t} + \sqrt{t_+} }\Biggr]^n 
\sum_{k=0}^\infty a^{(0)}_k z(t;t_0) ^k
\nonumber\\
 &\equiv& \Bigl( \sum_j c_j^{(n)} z(t;t_0)^j \Bigr)
     \sum_{k=0}^\infty a^{(0)}_k z(t;t_0)^k
\ \ \ \  .
\end{eqnarray}
It is a simple matter to compute the coefficients $c_j^{(n)}$
and match powers of $z$.
The condition $ \sum \left( a^{(n)}_k \right)^2 < 1$  then implies
\begin{eqnarray}\label{newbd}
  (c_0^{(n)} a_0^{(0)} )^2 
   + (c_0^{(n)} a_1^{(0)} + c_1^{(n)} a_0^{(0)} )^2
   + (c_0^{(n)} a_2^{(0)} + c_1^{(n)} a_1^{(0)} + c_2^{(n)} a_0^{(0)}  )^2
    + \ldots < 1
\ \ \ \ ,
\end{eqnarray}
for each $n$. 
For a given set of $c_j^{(n)}$ one may find that the higher moments
provide tighter constraints on the $a_j^{(0)}$ than the lowest moment alone.
This is the case for $B\rightarrow \pi$,
for the lowest few moments.  However, 
for $B\rightarrow D$ the higher moments do not improve 
the bounds imposed by the lowest moment.

\bigskip
\bigskip
\centerline{\bf Constraints for $\bar B \to \pi  l \overline \nu$}
\bigskip
\bigskip

To make use of Eq.~(\ref{nmaster})\ 
in extracting $V_{ub}$ from experimental measurements of  
$ \bar B \to \pi  l \overline \nu$ 
we should keep only a finite number of parameters $a_k^{(0)}$ and
compute the maximal truncation error from the omission of 
higher order terms in
the series. This truncation error can be minimized by
optimizing $t_0$ and thereby decreasing $z_{max}$\cite{bl}.
Applying Eq.~(\ref{newbd})\ further decreases the
truncation error by restricting the range of higher order
parameters (of course, it also restricts the range of 
the parameters we keep). 
Both effects decrease the number
of parameters required to determine $f_+(t)$ at a given level
of accuracy.

Experimental data on $\bar B \to \pi l \overline \nu$ is
not yet available to precisely describe the shape of $f_+(t)$ and 
so for now we will simply illustrate the utility
of the  higher moments. Imagine
$f_+(t)$ is known at some fixed number of kinematic 
points. For concreteness we choose the 
lattice-inspired values\cite{lellouch,ukqcd,ape}
$f_+\bigl(21\, {\rm GeV}^2\bigr) = 1.7$ and 
$f_+\bigl( 0 \, {\rm GeV}^2\bigr) = 0.5$
to fix $a_0^{(0)}$ and $a_1^{(0)}$ in terms of $a_2^{(0)}$.
Varying $a_2^{(0)}$ subject to the zeroth moment constraint
$\sum (a_k^{(0)})^2 < 1$ maps out the 
envelope of parameterizations
\footnote{
This procedure
is equivalent to the method of computing upper and lower bounds
by forming determinants of inner products often used in the
literature\cite{hist,brm,bgl1,lellouch,becirevic}. 
However,
Eq.~(\ref{nmaster})\ 
also dictates the  shape of curves 
allowed inside the envelope.
}
consistent with the $n=0$ dispersion relation. 
Varying
$a_2^{(0)}$ subject to the constraint of the 
$n^{th}$ moment results in a smaller allowed range for
$a_2^{(0)}$, for the first few $n$. 
Since the allowed range of $f_+(t)$
is proportional to the allowed range of $a_2^{(0)}$, 
a relative reduction in the range of $a_2^{(0)}$ leads 
to the same relative
reduction in the width of the envelope.

The one-loop results for the allowed range of $a_2^{(0)}$
are shown in the second column of Table~1. We have used
a pole quark mass $m_b = M_B - \bar \Lambda$ corresponding
to $\bar \Lambda = 0.4 $\ GeV\cite{lbar} (varying
$\bar \Lambda$ by $\pm 0.1 $\ GeV results in no more
than a $6\%$ change in the bounds), and fixed 
$t_0=t_-$.  
Note that the $n=3$ bounds are tighter than the $n=0$ result by
a factor of three. In fact, the $n=6$ bound (not listed) is better 
by a factor of four. However, $\chi^{(n)}$ receives known\cite{generalis}
corrections
from two-loop perturbative graphs (${\cal O}\left(\alpha_s\right)$)
and nonperturbative matrix elements. We use values of the condensates 
$< \bar u u> |_{1 {\rm GeV}} = (-0.24\ {\rm GeV})^3$ and
${\alpha_s\over \pi} <G_{\mu\nu}G^{\mu\nu}> = 0.02\pm 0.02\ {\rm GeV}^4$
from reference \cite{ball}.  The third 
and fourth columns of Table 1 show the perturbative and condensate
correction factors $ \rho^{(n)}_{pert}$ and $\rho^{(n)}_{cond}$ defined by
\begin{eqnarray}\label{rhon}
\chi^{(n)}(two \,loop) &=& (1 + \rho^{(n)}_{pert} 
+ \rho^{(n)}_{cond})\chi^{(n)}(one \,loop)
\ \ \ \ .
\end{eqnarray}
Beyond $n=3$ the sum of these corrections approaches $100\%$ of the 
leading result and the reliability of the calculation becomes
questionable. Even at $n=3$, the corrections are large enough that 
one might worry about the convergence of the operator product
expansion. We will use $n=2$ in our examples.
The allowed range of $a_2^{(0)}$ using
the two-loop result, including condensates, is shown in the fifth 
column of Table~1. 
Although weaker than the corresponding one-loop result, the
two-loop bounds for $n > 0$ remain  
significantly more constraining than the  $n=0$ bounds.

\bigskip
\vbox{\medskip
\hfil\vbox{\offinterlineskip
\hrule
\halign{&\vrule#&\strut\quad\hfil$#$\quad\cr
height2pt&\omit&&\omit&&\omit&&\omit&&\omit&&\omit&\cr
&n && a_2^{0}(one \,loop) && \rho^{(n)}_{pert} && \rho^{(n)}_{cond}
&& a_2^{0}(two \,loop)&& a_2^{(0)}(HQS)& \cr
\noalign{\hrule}
& 0 && -0.77 < a_2 < 0.79 && 0.23 
&& 0.01 && -0.78 < a_2 < 0.79 && -0.54< a_2< 0.58 &\cr
&1 &&  -0.45 < a_2 < 0.47
&& 0.32 &&
0.06 && -0.52< a_2 < 0.53 &&-0.40 < a_2 < 0.41 &\cr
&2&&  -0.33 < a_2 < 0.34 && 0.38
&& 0.14 &&-0.44 < a_2 < 0.45 && -0.36 < a_2 < 0.36 &\cr
&3 && -0.26 < a_2 < 0.28 && 0.42 
&& 0.28 &&-0.40< a_2 < 0.41 && -0.35 < a_2 < 0.35&\cr }
\hrule}
\hfil}
\medskip
\INSERTCAP{1}{Values of the zeroth moment parameter $a_2^{(0)}$ 
consistent with the $n^{th}$ moment constraint.  From 
left to right is the constraint on $a_2^{(0)}$ at one loop,
the relative two-loop perturbative and condensate contributions 
$\rho^{(n)}_{pert}$ and $\rho^{(n)}_{cond}$, the constraint on $a_2^{(0)}$ at
two loops, and the two-loop constraint on $a_2^{(0)}$ 
including the $B^*\pi$ spin symmetry contribution
as described in the text.
}

The ellipsis in Eq.~(\ref{optical})
includes the contribution of intermediate 
$\bar B^* \pi$ states to the hadronic moments. 
In the heavy b-quark limit, the form factor $g_*$ defined by 
\begin{eqnarray}\label{gstar}
  \bra{\pi(p')} V^\mu \ket{\bar B^*(p,\epsilon)} &=&
\, i g_* \epsilon^{\mu \alpha \beta \gamma} \epsilon_\alpha \,
p_{\beta}' \, p_{\gamma}
\ \ \ \ ,
\end{eqnarray}
is related to $f_+$ by spin symmetry when the $\pi$ meson is soft
({\it i.e.} near $t = t_-$),
\begin{eqnarray}\label{hqsrelation}
g_*(v\cdot p') = {2\over M_B} f_+(v\cdot p') [ 1 + {\cal O}(\frac1{M})]
\ \ \ \ ,
\end{eqnarray}
where $v = p/M_B$. 
Including the contribution of the 
$\bar B^* \pi$ state (similar contributions 
were used in \cite{caprinim}\ for the $B \to B$ elastic form factor)
modifies the constraint of  Eq.~(\ref{Ptint})\ to
\begin{eqnarray}\label{Ptint2}
{1\over \pi}\int_{t_+}^\infty dt 
\left[ 
 |P(t) X(t) \phi^{(n)}_{f_+}\  f_+(t)|^2 +  |P(t) X(t) \phi^{(n)}_{g_*}\  g_*(t)|^2 
 \right] \le 1
 \ \ \ \ ,
\end{eqnarray}
where $X(t) =  [ (t_+ -t_0)/ (t_+ -t)]^{1/4} 
        / (\sqrt{t_+ -t} + \sqrt{t_+ -t_0} ) $.
One finds that $\phi_{g_*} = \phi_g$, so expanding
\begin{eqnarray}\label{nmaster2}
g_*(t) = {1\over P(t) \phi^{(n)}_g(t)}
\sum_{k=0}^\infty b^{(n)}_k z(t;t_0)^k
\ \ \  ,
\end{eqnarray}
leads to the constraint
\begin{eqnarray}\label{bsum}
\sum_{k=0}^\infty ( a_k^{(n)} )^2 +  ( b_k^{(n)} )^2 \le 1
\ \ \ \ .
\end{eqnarray}
We cannot relate all of the $b_k^{(n)}$ to the
$a_k^{(n)}$ because the b-quark spin symmetry is only
valid near $t = t_-$. 
One expects that the normalization and first derivative of
$g_*$ and $f_+$ at $t_-$ obey the heavy quark relation
to $\sim 10\%$ for the physical B mass, so 
\begin{eqnarray}\label{b0b1}
b_0^{(n)} &=& \frac2{M_B} { \phi^{(n)}_g(t_-) \over \phi^{(n)}_{f_+}(t_-)} a_0^{(n)}
\nonumber\\
&=& \sqrt{2} \left(1 + \sqrt{\frac{M_\pi}{M_B}}\right)^2 a_0^{(n)}
\ \ \  ,
\nonumber\\
b_1^{(n)} &=& 4 \sqrt{2} \sqrt{\frac{M_\pi}{M_B}} a_0^{(n)} 
+  \sqrt{2} \left(1 + \sqrt{\frac{M_\pi}{M_B}}\right)^2 a_1^{(n)} 
\ \ \ \  ,
\end{eqnarray}
to $\pm 10\%$. 
Taking $90\%$ of the absolute values of the right-hand sides of Eq.~\ref{b0b1} 
gives lower bounds on $|b_0^{(n)}|$ and $|b_1^{(n)}|$, that can
be inserted into Eq.~(\ref{bsum}) (this constraint applies as well to heavy-heavy
systems, where it could be further improved by including the $\bar B^* D^*$
intermediate state). 
The improvement on the range of $a_2^{(0)}$ from the inclusion of 
this additional hadronic final state  is
shown in the last column of Table 1 
(using two-loop amplitudes). 
The effect is $\sim 30\%$  for $n=0$, decreasing to 
$\sim 15\%$ for $n=3$. 
\begin{figure}
\epsfxsize=16cm
\hfil\epsfbox{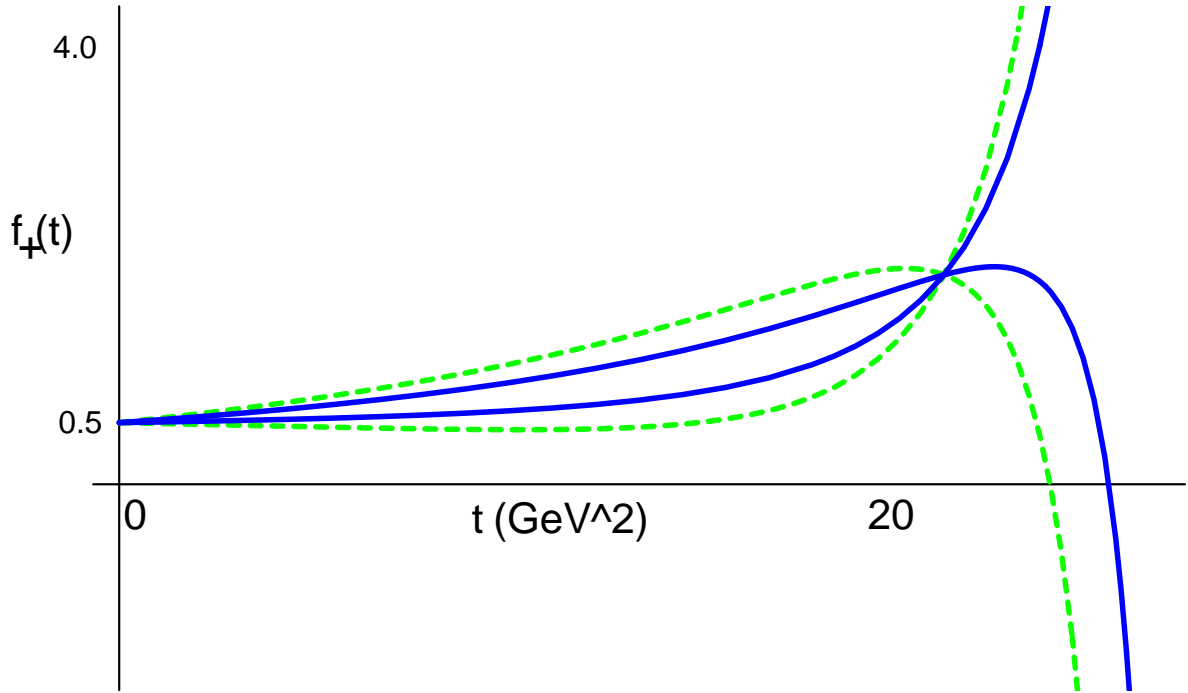}\hfill
\caption{ Upper and lower bounds on the
$\bar B \to \pi l \overline \nu $ form factor $f_+(t)$.
The Dashed lines are the bounds arising from the $n=0$ moment
while the  solid lines are the bounds arising from the
$n = 2$ moment.  The form factor has been fixed at
two points by the lattice inspired values
$f_+(21 {\rm GeV}^2) = 1.7$ and $f_+(0) = 0.5$.
The allowed regions are the interiors of the dashed or
solid pairs of curves.}
\end{figure}

The  $n=0$ bounds on $f_+(t)$ at two loops, without the 
application of spin symmetry (column five of Table~1), are
shown as the outer, dashed, pair of curves in Fig.~2,  
while the bounds from the $n=2$ improvement, 
including the contribution from $\bar B^* \pi$, 
are given by the inner, solid, pair of curves.
The constraints arising from the $n\ge 0$ 
moments are a dramatic improvement over the 
lowest order $n=0$ constraint alone. Note that
even the $n=1$ bounds in the last column of Table~1
represent an improvement over that of $n=0$ alone
by nearly a factor of two.
\begin{figure}
\epsfxsize=16cm
\hfil\epsfbox{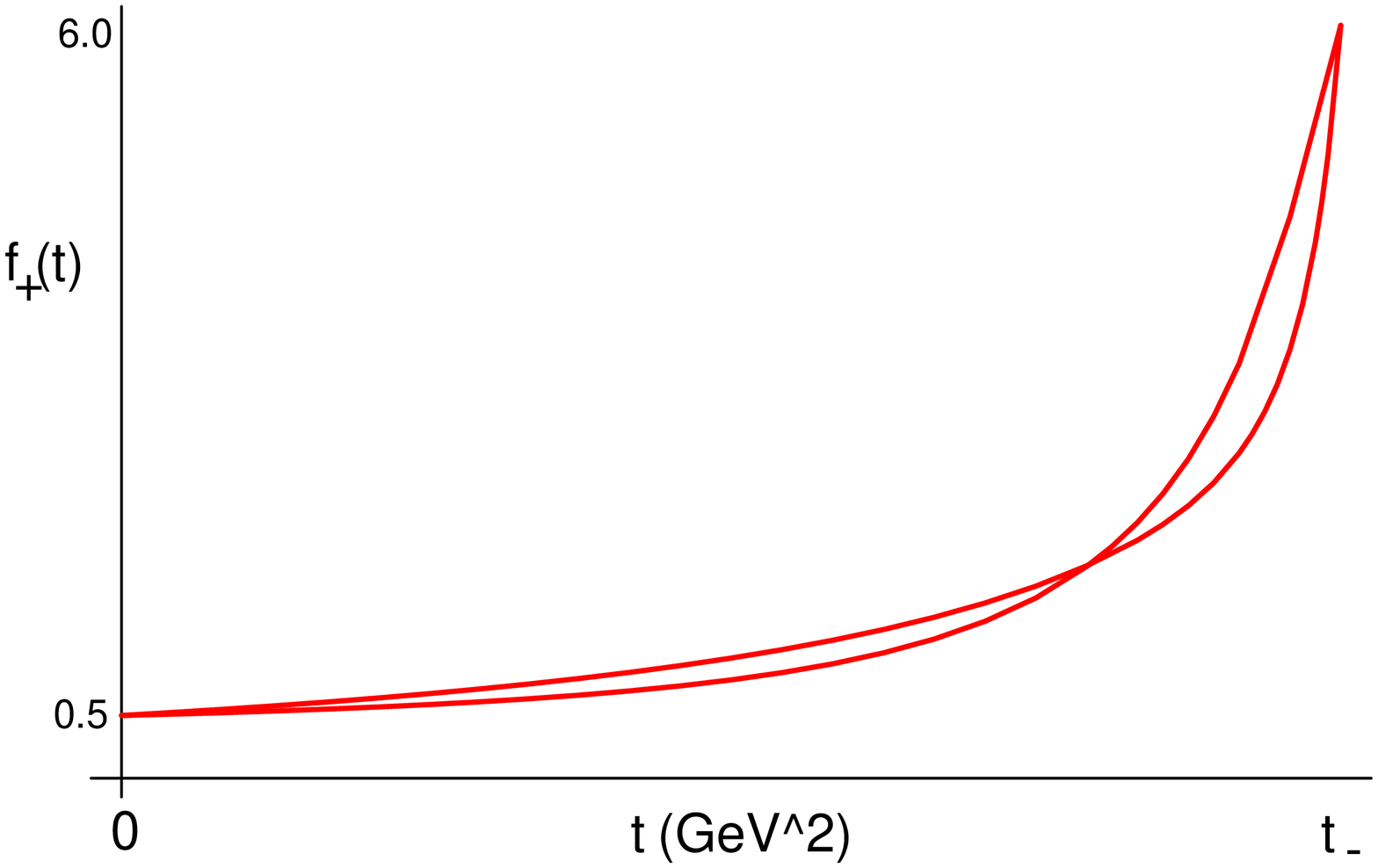}\hfill
\caption{ Bounds on the
$\bar B \to \pi l \overline \nu $ form factor $f_+(t)$
arising from three normalization points,
$f_+( t_- ) = 6.0$,\
$f_+(21\  {\rm GeV}^2) = 1.7$ and
$f_+(0) = 0.5$.
The constraints are from
moments of $n \le 2$ as discussed in the text, including
the $B^*\pi$ hadronic intermediate state using bottom-quark spin symmetry
at $t=t_-$.
}
\end{figure}

The quickly weakening constraint near $t = t_-$ would be absent had
we included a normalization at zero recoil.  Fig.~3 shows the spin-improved,
$n=2$ constraint assuming an additional normalization, $f_+(t_-)=6$.
It gives a feel for how tightly $f_+(t)$ can be constrained given
measurements of $f_+ (t)$ at only three points.
In this case, $f_+(t)$ is determined
to better than $\pm 15\%$ over the entire kinematic region. 
These plots are presented only to provide 
a feel for what is needed to describe
$f_+(t)$ to a given accuracy. 
In practice, one should fit the parameterization
Eq.~(\ref{nmaster})\ to experimental data and 
theoretical (lattice or heavy quark
symmetry) predictions to extract $V_{ub}$. 
The work presented here 
should improve the precision of such extractions significantly.

\bigskip
\bigskip
\bigskip
\bigskip
\centerline{\bf Conclusions}
\bigskip
\bigskip

We have discussed constraints on semileptonic form factors
arising from analyticity and dispersion relations, and 
identified the physical origin of the naturally small
parameter $z_{max}$. The  shape and magnitude of an
analytically continued form factor in the pair-production region
has significant impact upon the size and shape of 
the form factor in the semileptonic region. 
Perturbative constraints
on pair-production  therefore  constrain semileptonic
decay amplitudes. 
Such constraints take the form of parameterizations of
form factors in terms of a small number of unknown, but bounded,
constants $a_k$. 
Only a small number of  $a_k$ are needed to
describe a given form factor because the parameterizations arise
from truncated expansions in a kinematic variable $z(t;t_0)$
that is surprisingly small, $ | z(t;t_0)| < z_{max} = 0.065$ for \btods\
and $ | z(t;t_0)| < z_{max} = 0.52 $ for $\bar B \to \pi l \overline\nu$.

We have traced the smallness of $z_{max}$ to the dual nature of the
hadronic and partonic descriptions of QCD. Whereas in the hadronic
description the form factor is a sum of poles and less singular terms, 
in the partonic description the form factor is always less singular than a 
pole. Duality then implies that the variation of
the form factor $F(t)$ over the physical region 
$ 0< t < t_-$ for the semileptonic decay $\bar B \to H l \bar \nu$ is
characterized by
\begin{eqnarray}
{ F(t_-) - F(0) \over F(t_-) +  F(0) } \sim
 \left( {\sqrt{M_B}-\sqrt{M_H}\over \sqrt{M_B}+\sqrt{M_H} }\right)^2
= z_{max} \ \  .
\end{eqnarray}
The small values of $z_{max}$ are due to the square-root dependence
on meson masses, which is in turn a consequence of the absence
of poles in the partonic description of QCD.

We applied these ideas in a discussion of
analyticity constraints in terms of the momentum-transfer
variable $t$. 
We improved the constraints on the $f_+ (t)$ form factor 
by $\sim 15-30\%$
by including the contribution 
of the $\bar B^* \to \pi l \bar \nu$ form factor using 
b-quark spin symmetry to relate it to
the $\bar B \to \pi l \bar \nu$ form factor at zero recoil.
We also improved the analyticity constraints by 
considering higher moments of the dispersion relation. 
As an illustrative example, we
fixed the normalization of the $\bar B \to \pi l \bar \nu$
form factor $f_+(t)$ at two kinematic points to lattice-inspired
values, and plotted the envelope of allowed parameterizations (Fig.2).
The higher moment constraints result in upper and lower bounds
that are roughly twice as tight as the traditional, 
zeroth-moment result. Given a third normalization at
zero recoil, the form factor is determined over the
entire kinematic range to $\pm 15\%$, as shown in Fig.3. 
This suggests that
a parameterization using one normalization (possibly given
by lattice or heavy quark symmetry predictions)  and two free 
parameters could eventually be used for a model-independent extraction
of $V_{ub}$ with roughly $30\%$ theoretical errors arising from the
$t$ dependence.

\vskip 1.5cm

\centerline{\bf Acknowledgments}
\bigskip
\bigskip

We would like to thank B. Grinstein, M. Lu, and M. Wise for useful discussions.
This work is supported by the Department of Energy under Grant
no. DE-FG02-91-ER 40682. 
CGB thanks the Nuclear Theory group 
at the University of Washington for their hospitality during
part of this work.

\end{document}